\documentclass[preprint,aps,showpacs]{revtex4}

\usepackage{amssymb} \usepackage{graphicx}
\usepackage{dcolumn} \usepackage{amsmath}

\begin{document}

\title{Deformed Clifford algebra and supersymmetric quantum mechanics on a phase space
with applications in quantum optics}

\author{{\. I}. Bu{\u g}dayc{\i}}
\email{bugdayci@science.ankara.edu.tr}
\address{Department of Physics, Ankara University, Faculty of Sciences, 06100, Tando{\u g}an-Ankara, Turkey}

\author{A. Ver{\c c}in}
\email{vercin@science.ankara.edu.tr}
\address{Department of Physics, Ankara University, Faculty of Sciences, 06100, Tando{\u g}an-Ankara, Turkey}


\date{\today}
\begin{abstract}

In order to realize supersymmetric quantum mechanics methods on a
four dimensional classical phase space, the complexified Clifford
algebra of this space is extended by deforming it with the Moyal
star-product in composing the components of Clifford forms. Two
isospectral matrix Hamiltonians having a common bosonic part but
different fermionic parts depending on four real-valued phase space
functions are obtained. The Hamiltonians are doubly intertwined via
matrix-valued functions which are divisors of zero in the resulting
Moyal-Clifford algebra. Two illustrative examples corresponding to
Jaynes-Cummings-type models of quantum optics are presented as
special cases of the method. Their spectra, eigenspinors and Wigner
functions as well as their constants of motion are also obtained
within the autonomous framework of deformation quantization.\

\textbf{Keywords:} Clifford algebra, supersymmetric quantum
mechanics, deformation quantization, Wigner functions,
Jaynes-Cummings models
\end{abstract}

\pacs{45.20.Jj, 11.30.Pb, 03.65.Fd}

\maketitle

\section{Introduction}

Moyal star-product formalism, more generally deformation
quantization, provides useful tools for describing the phase space
formulation of quantum mechanics \cite{Bayen,Zachos2}. Spin and
relativistic quantum mechanics can also be considered in this
framework by making use of an additional fermionic star product
\cite{Henselder1}. It is seen that a combination of the Clifford
product as fermionic part and Moyal star-product as bosonic part
plays an important role in these regards. This approach can also be
adopted to realize the methods of supersymmetric quantum mechanics
\cite{Gendensthein,Junker,Cooper,Kuru} on a phase space. Henselder
has already shown how such an approach can be accomplished on a
two-dimensional ($2D$) phase space in a recent study
\cite{Henselder2}. The main goal of the present work is to show that
such a programme can be realized on a $4D$ phase space in its full
generality and rich applicability potential.

In this paper, we first study the deformation of the complex
Clifford algebra defined on a $4D$ phase space of a classical
Hamiltonian system. The deformation is performed in composing the
components of Clifford forms with the Moyal star-product. This
enables us to obtain two iso-spectral matrix Hamiltonians, each
being a sum of factorized terms. The Hamiltonians depend on four
real-valued phase-space functions and they are general enough for
modeling various systems. Moreover, the Hamiltonians are also doubly
intertwined in such a way that the intertwining matrix-valued
phase-space functions are divisors of zero in the resulting deformed
algebra. By composing the intertwining matrices with their
conjugates, we obtain a constant of motion for each system.
Well-known Jaynes-Cummings-type models of quantum optics
\cite{JC1,JC2,Schleich}, which are the main subject of recent
experiments to understand the quantum nature of atom-field
interactions \cite{Nature1}, emerge as special cases of our method.
These are studied in detail as the illustrative examples. Their
phase-space characteristics such as their spectra, eigenspinors and
related Wigner functions as well as their classical limits and
constants of motion are obtained in an autonomous way, without any
reference to the standard tools of quantum mechanics.

The paper is structured into two main parts. The first part consists
of the following three sections where our method is fully developed.
In Sec. II the Clifford algebra structure of $4D$ classical phase
space and its complexification are studied and the basics of Moyal
star-product are briefly reviewed. The deformation of the algebra
and its application for establishment of supersymmetry techniques on
the phase space are taken up in Sec. III and its physical
consequences are studied in Sec. IV. Sec. V, which constitutes the
second part of the study, is entirely devoted to the applications of
the method in quantum optics. Concluding remarks are given in the
final section.

\section{Clifford Algebra on a Phase-Space}

Any vector space on which a non-degenerate inner product is defined
has a Clifford algebra structure. Tangent (or cotangent) space of
any (pseudo) Riemannian manifold is the best place where efficiency
of Clifford calculus can be observed \cite{Benn-Tucker}. In this
regard the symplectic manifold that we shall consider in this and
the following section can be any even-dimensional manifold. However,
in order to be able to do quantization consistently in such a
framework, one needs the star product of deformation quantization
(see subsection IIC) which requires globally defined coordinates.
For this reason and for physically relevant applications, we shall
confine the investigation to $4D$ flat cases. Our method can easily
be extended to an arbitrary flat symplectic manifold. For
definiteness, we take our phase space to be a $4D$ symplectic
manifold ${\cal M}$ topologically equivalent to $\mathbb{R}^{4}$ and
denote the linear space of complex-valued, smooth (differentiable to
all orders) functions defined on it by ${\cal F}$. Real-valued
elements of ${\cal F}$ are the classical observables of Hamiltonian
mechanics.

At each point of ${\cal M}$ two $4D$ vector spaces, the tangent
space and (its dual) the cotangent space are naturally defined.
Elements of the latter are $1$-forms acting linearly on the elements
of the former. ${\cal M}$ carries two distinguished structures
defined by two non-degenerate, second rank tensor fields: the
symplectic $2$-form $\Omega$ (which is also closed) and the
Euclidean symmetric metric tensor $g$. They are non-degenerate in
the sense that each defines a vector space isomorphism between the
tangent and cotangent spaces. The elements associated by these
isomorphisms are called symplectic and metric dual of each other.
The metric dual of a vector field $x$ is a $1$-form $\tilde{x}$
defined, for a given vector field $y$, as
$\tilde{x}(y)=i_y(\tilde{x})=g(x,y)$. Here $i_{y}$ denotes the
so-called interior derivation (or interior product) whose action on
an arbitrary $k$-form (a totally anti-symmetric covariant tensor)
$\beta$ is defined, for arbitrary vector fields $x_1,\dots x_{k-1}$,
by $(i_{y}\beta)(x_1,\dots, x_{k-1})=k\beta(y,x_1,\dots x_{k-1})$.
$\Omega$ and $g$ at each point of a given neighborhood endow the
tangent space with structures of a symplectic vector space and an
inner product space, respectively. According to the Darboux theorem
\cite{Arnold}, in each neighborhood of ${\cal M}$ one can define the
canonical coordinates $({\bf q},{\bf p})=(q_{1},q_{2},p_{1},p_{2})$
which lead to $\Omega=\sum_j dq_j\wedge dp_j$. Here $\wedge$ denotes
the bilinear and associative exterior product which satisfies
\begin{equation}
\alpha\wedge\beta=(-1)^{jk}\beta\wedge\alpha\;,\nonumber
\end{equation}
for arbitrary $j$-form  $\alpha$ and  $k$-form $\beta$. $\Omega$
eventually leads to the Poisson bracket (PB)
\begin{equation}
[F,G]_{P}=\sum_{k=1}^{2}(\partial _{q_{k}}F\partial
_{p_{k}}G-\partial _{p_{k}}F\partial _{q_{k}}G)\;
\end{equation}
of Hamiltonian mechanics. We henceforth use the abbreviation
$\partial _{q_{i}}\equiv \partial /\partial q_{i}$ and denote the
generic elements of ${\cal F}$ by capital Latin letters. On the
right-hand side of (1), the following ordinary commutative and
associative pointwise product of functions is essential:
\begin{equation}
(F_{1}F_{2})({\bf q},{\bf p})=(F_{2}F_{1})({\bf q},{\bf
p})=F_{1}({\bf q},{\bf p}) F_{2}({\bf q},{\bf p})=F_{2}({\bf q},{\bf
p}) F_{1}({\bf q},{\bf p})\;.\nonumber
\end{equation}

\subsection{Clifford Bundle and Complexification}

The linear space of all forms constitutes the exterior algebra with
respect to exterior product, which (like the tensor algebra) is a
$\mathbb{Z}$-graded associative algebra. Although, unlike the tensor
algebra, the exterior algebra is finite dimensional ($2^{n}D$ when
the dimension of ${\cal M}$ is $n$) its $\mathbb{Z}$-gradation is
inherited from the tensor algebra. In particular, any $k$-form for
$k>n$ is zero and the zero element of the exterior algebra is the
only homogeneous element of every degree greater than $n$ (see also
\cite{Benn-Tucker} pp. 7-8).

The Clifford product $\ast_{C}$, defined for a $1$-form $\tilde{x}$
and an arbitrary form $\beta$ by
\begin{equation}
\tilde{x}\ast_{C} \beta = \tilde{x} \wedge \beta + i_x \beta\;,
\label{}
\end{equation}
turns the exterior algebra into a $\mathbb{Z}_2$-graded associative
algebra called the Clifford algebra (also known as the geometric
algebra \cite{Artin,Hestenes1}. See also \cite{Hestenes2} for
applications in Hamiltonian mechanics). By associativity, the rule
(2) suffices to completely determine $\ast_{C}$ on arbitrary forms.
$\mathbb{Z}_2$-gradation means that the whole algebra is, as a
linear space, a direct sum of the spaces of the odd and even forms
such that the latter has a subalgebra structure. The Clifford
product of a $j$-form $\alpha$ and $k$-form $\beta$ is, in general,
an inhomogeneous form consisting of a sum of $\ell$-forms such that
$\ell=j+k,j+k-2,\dots, |j-k|$. The exterior bundle (union of
exterior algebras) equipped with the $\ast_{C}$-product in the
fibres is called the Clifford bundle. The Clifford commutator,
defined by
\begin{eqnarray}
[\alpha, \beta]_{C} = \alpha \ast_{C} \beta - \beta \ast_{C} \alpha
\;,\nonumber
\end{eqnarray}
is bilinear, antisymmetric and satisfies the Jacobi identity. Each
fibre of the Clifford bundle acquires a Lie algebra structure with
this bracket. In the context of Clifford algebra, we shall mainly
adopt the conventions of \cite{Benn-Tucker}, except for the
representation of the Clifford product. More conventional versions
of this product are $\tilde{x}\ast_{C}\beta =\tilde{x}\vee\beta$ and
juxtaposing the factors.

Tangent spaces can be complexified by replacing the field of real
numbers $\mathbb{R}$ by the field of complex numbers $\mathbb{C}$.
Then $g$ can be extended by $\mathbb{C}$-linearity to the
complex-valued, symmetric and non-degenerate (guaranteed by the
non-degeneracy of $g$) bilinear map $g^{\mathbb{C}}$.  As
$\mathbb{C}$ is algebraically closed, $g^{\mathbb{C}}$ is not
characterized by any signature (even if $g$ had any non-Euclidean
signature) and hence the structure of complex algebra depends only
on the dimension. In our case, we consider $2^{4}D$ real Clifford
algebra $\mathbf{ C}_{4,0}(\mathbb{R})$ which is isomorphic to the
algebra of $2\times 2$ quaternion matrices (\cite{Benn-Tucker},
p.80). But from here on, we shall deal with its complexification
$\mathbf{C}_4(\mathbb{C})$ which is known to be isomorphic to the
algebra of $4\times 4$ complex matrices.

\subsection{The Primary SUSY Structure}

The $2^{4}D$ complex algebra $\mathbf{C}_4(\mathbb{C})$ is generated
by $1$ and by the complex orthonormal basis $\{e^{1}, e^{2}, e^{3},
e^{4}\}$ such that
\begin{eqnarray}
e^{j}\ast_{C}e^{k}+ e^{k}\ast_{C}e^{j}=2\delta^{jk}\;,
\end{eqnarray}
where the Kronecker symbols denotes the components of the (inverse)
metric $g^{\mathbb{C}}(e^{j},e^{k})=\delta^{jk}$. We start out our
analysis by the Clifford $1$-form
\begin{eqnarray}
\omega = W_{1} e^{1} + W_{2} e^{2} + P_{1} e^{3} + P_{2} e^{4}\;,
\end{eqnarray}
whose components are real-valued phase-space functions
\begin{eqnarray}
W_{j} = W_{j}(\mathbf{q}, \mathbf{p}),\;P_{j} = P_{j}(\mathbf{q},
\mathbf{p})\;, \; j=1,2. \nonumber
\end{eqnarray}
A rather general Hamiltonian function in two dimensions can now be
written, in terms of $\omega $, as a Clifford product
\begin{eqnarray}
H=\frac{1}{2}{\mathbf{\omega}}*_{C}\omega=
\frac{1}{2}(P_{1}^2+P_{2}^2)+\frac{1}{2}({W_1}^2+{W_2}^2)\;.
\end{eqnarray}
We next introduce the $1$-forms
\begin{eqnarray}
f&=&\frac{1}{\sqrt{2}}(e^{1}+ie^{3})\;,
\quad \check{f}=\frac{1}{\sqrt{2}}(e^{1}-ie^{3}),\nonumber\\
g&=&\frac{1}{\sqrt{2}}(e^{2}+ie^{4})\;, \quad
\check{g}=\frac{1}{\sqrt{2}}(e^{2}-ie^{4}).\nonumber
\end{eqnarray}
These are all nilpotent of order 2
\begin{eqnarray}
f*_C f=0=g*_C g\;,\quad \check{f}*_C \check{f}= &0&=\check{g}*_C
\check{g}\;,
\end{eqnarray}
and satisfy
\begin{eqnarray}
\;\{f,\check{f}\}_C=& 2 &=\{g,\check{g}\}_C\;,\nonumber\\
\;\{g,\check{f}\}_C=&0&=\{\check{g},f\}_C\;,\\
\;\{g,f\}_C=&0&=\{\check{g},\check{f}\}_C\;,\nonumber
\end{eqnarray}
where $\{,\}_C$ denotes the Clifford anti-commutator. The relations
(6) and (7) imply that the set
$\{2^{-1/2}f,2^{-1/2}g,2^{-1/2}\check{f},2^{-1/2}\check{g}\}$
constitutes a Witt basis of the complexified cotangent space. The
first two and the last two elements of this basis span two isotropic
subspaces (where $g^{\mathbb{C}}$ induces the zero bilinear map)
whose direct sum is the whole cotangent space. Then in terms of the
complex-valued functions
\begin{eqnarray}
C_1=\frac{1}{\sqrt{2}}(W_{1}+iP_{1})\;,\quad
C_2=\frac{1}{\sqrt{2}}(W_{2}+iP_{2})\;,
\end{eqnarray}
and their complex conjugates $\bar{C}_1,\bar{C}_2$, we define
\begin{equation}
q_{-}=\bar{C}_1f+\bar{C}_2g\;,\quad
q_{+}=C_1\check{f}+C_2\check{g}\;,
\end{equation}
such that (4) can be rewritten as $\mathbf{\omega}=q_{+}+q_{-}$. By
using (6) and (7), one can easily verify that $q_{\pm}$ are also
nilpotent, and together with $H$, they close in a simple
supersymmetric algebra structure
\begin{eqnarray}
q_{\pm}*_C q_{\pm}&=&0\;,\nonumber\\
H&=&\frac{1}{2}\{q_{-},\;q_{+}\}_{C}\;,\\
\;[q_{\pm},H]_C&=&0\;.\nonumber
\end{eqnarray}

The Clifford algebra structure of the exterior bundle of ${\cal M}$
enabled us to see the supersymmetry (SUSY) structure in the
corresponding classical system. Such systems possessing the
fermionic (anti-commuting elements) degrees of freedom, in addition
to the usual bosonic ones, are known as pseudoclassical models (see
\cite{Junker} for an extensive list of references). They serve as
the classical limits of quantum systems having both kinds of degrees
of freedom. However, the SUSY structure emerging above does not seem
to be promising  much at this stage since the third relation of (10)
is trivially satisfied. Indeed, $H$ is a $0$-form and therefore
Clifford commutes with all forms. In general, the third relation of
(10) must be a consequence of the nilpotent nature of so called
supercharges $q_{\pm}$. In what follows, the algebra will be
deformed in such a way that the last three relations with a new
Hamiltonian, which turns out to be an inhomogeneous even Clifford
form comprising a $0$-form and a $2$-form, will manifest a genuine
SUSY structure. For this purpose, the basics of Moyal star-product
and its corresponding bracket are briefly reviewed in the following
subsection.

\subsection{Star Product and Moyal Bracket}

In the canonical $({\bf q},{\bf p})$ coordinates, the Moyal $\ast
$-product on ${\cal F}$ is defined by
\begin{equation}
\ast =\exp \Big[\frac{1}{2}i\hbar \sum_{j=1}^{2}\Big(\stackrel{\leftarrow }{%
\partial }_{q_{j}}\stackrel{\rightarrow }{\partial }_{p_{j}}-\stackrel{%
\leftarrow }{\partial }_{p_{j}}\stackrel{\rightarrow }{\partial
}_{q_{j}}\Big)\Big],
\end{equation}
where $\hbar$ is the Planck constant and $\stackrel{\leftarrow
}{\partial },\;\stackrel{\rightarrow }{\partial }$ are acting,
respectively, on the left and on the right. This product is
bilinear, associative, and obeys the relation
\begin{equation}
(\overline{F_{1}\ast F_{2}})=\bar{F}_{2}\ast \bar{F}_{1}\;,
\end{equation}
under complex conjugation. In terms of the $\ast$-product the Moyal
bracket $[,]_{M}$ is defined as
\begin{equation}
[F,G]_{M}=F\ast G-G\ast F\;,
\end{equation}
for all phase-space functions. Note that in view of (12), we have
\begin{equation}
\overline{[F,G]}_{M}=-[\bar{F},\bar{G}]_{M}\;.
\end{equation}
In particular, the Moyal bracket of two real-valued functions is a
purely imaginary-valued function. The most important properties of
the $\ast $-product and the MB are the following limiting relations
\begin{eqnarray}
\lim_{\hbar \rightarrow 0}F\ast G&=&FG\;,\nonumber\\
\lim_{\hbar \rightarrow 0}\frac
{1}{i\hbar}[F,G]_{M}&=&[F,G]_{P}\;,\nonumber
\end{eqnarray}
which hold for generic $\hbar$-independent phase-space functions.
These reveal the fact that the associative $\ast$-algebra and the
Lie algebra structure of ${\cal F}$ given by the MB are,
respectively, deformations (in the sense of Gerstanhaber
\cite{Gerstenhaber}) of associative algebra structure of ${\cal F}$
with respect to pointwise product and of Lie algebra structure
determined with respect to PB. All quantum effects are encoded in
the $\ast$-product with respect to which the real elements of ${\cal
F}$ are promoted to the status of quantum observables.

\section{Moyal-Clifford Algebra}

So far the components of Clifford forms were commuting quantities
since they were multiplied by the ordinary pointwise product.
However, one can go over to the non-commutative or quantum case by
demanding that the coefficients are to be multiplied by the Moyal
$*$-product. This can be achieved by combining (2) and (11)
together. The resulting associative product and algebra will be
referred to as the Moyal-Clifford (MC) product, denoted by $*_{MC}$
and as the MC-algebra, respectively. We will directly apply this
product to the above formulation. In doing so, we would like the
first relation of (10) to remain intact with respect to this new
product as well.

\subsection{SUSY structure by deformation}

It is easy to show that
\begin{eqnarray}
q_{+}*_{MC} q_{+}&=&[C_1,C_2]_{M}\check{f}*_{C}\check{g}\;,\nonumber\\
q_{-}*_{MC}
q_{-}&=&[\bar{C}_1,\bar{C}_2]_{M}f*_{C}g=-\overline{[C_1,C_2]}_{M}f*_{C}g\;,\nonumber
\end{eqnarray}
where in the last equality we have used (14). Thus $q_{+}$ and
$q_{-}$ are nilpotent with respect to the MC-product if and only if
\begin{eqnarray}
[C_1,C_2]_{M}=0\;.
\end{eqnarray}
In terms of $W_{j}$'s and $P_j$'s, this condition amounts to
\begin{eqnarray}
[W_1,W_2]_{M}-[P_1,P_2]_{M}=i[W_2,P_1]_{M}-i[W_1,P_2]_{M}.
\end{eqnarray}
Since $W_j$'s and $P_j$'s are real-valued, the left hand side of
(16) is, in view of (14), purely imaginary, while the right hand
side is real. So the condition (16) is, in fact, equivalent to the
following two conditions:
\begin{eqnarray}
\;[W_{1},W_{2}]_{M}&=&[P_1,P_2]_{M},\;\\
\;[W_{1},P_{2}]_{M}&=&[W_{2},P_{1}]_{M}\;.
\end{eqnarray}

We now define the SUSY Hamiltonian as
\begin{eqnarray}
H_{s}=\frac{1}{2}\{q_{+},q_{-}\}_{MC}\;,
\end{eqnarray}
which implies that
\begin{eqnarray}
[q_{\pm},H_s]_{MC}=0\;.
\end{eqnarray}
In terms of $\omega_{1}=\omega$ and $\omega_{2}=-i(q_{+}-q_{-})$
which obey $\{\omega_{1},\omega_{2}\}_{MC}=0$, $H_{s}$ can be
factorized as
\begin{eqnarray}
H_{s}&=&\frac{1}{2}\omega_{1}\ast_{MC}\omega_{1}=
\frac{1}{2}\omega_{2}\ast_{MC}\omega_{2}\;.
\end{eqnarray}
As will be shown in the following subsection, in an appropriate
matrix representation of the Clifford algebra, $\omega_j$ are
Hermitian supercharges and the symmetry determined by the above
super-algebra is $2$-extended supersymmetry denoted also by $N=2$
SUSY (where $N$ stands for the number of Hermitian supercharges).

Using (9), (19), conditions (17), (18) and
\begin{eqnarray}
\;[\bar{C_1},C_1]_M&=&i[W_1,P_1]_M\;,\quad
[\bar{C_2},C_2]_M=i[W_2,P_2]_M\;,\nonumber\\
\;[\bar{C_1},C_2]_M&=&[W_1,W_2]_M-i[W_2,P_1]_M\;,\nonumber\\
\;[\bar{C_2},C_1]_M&=&-[W_1,W_2]_M-i[W_2,P_1]_M\;,\nonumber
\end{eqnarray}
$H_s$ can explicitly be evaluated as
\begin{eqnarray}
H_{s}&=&H_{\ast}+\frac{1}{2}\big\{[W_{1},P_{1}]_{M}e^{13}+[W_{2},P_{2}]_{M}e^{24}\nonumber\\
 & &+[W_{2},P_{1}]_{M}(e^{14}+e^{23})+[W_{1},W_{2}]_{M}(e^{12}+e^{34})\big\}\;,
\end{eqnarray}
where $H_{\ast}$ should be read as $H_{\ast}1$ with $1$ being the
unit element of the Clifford algebra and
\begin{eqnarray}
2H_{\ast}&=&\{\bar{C_1},C_1\}_M+\{\bar{C_2},C_2\}_M\nonumber\\
&=&P_{1}\ast P_1+P_{2}\ast P_2+W_1*W_1+W_2*W_2\;.
\end{eqnarray}
Here $\{,\}_M$ stands for the anti-Moyal bracket and we have adopted
the abbreviation $e^{jk}=e^{j}\ast_{C} e^{k}$ which for the product
of different orthonormal basis elements becomes $e^{12}=e^{1}\wedge
e^{2}$, etc. Note that $H_s$ is an inhomogeneous even Clifford form
whose bosonic part $H_{\ast}$ is a zero form and the remaining
fermionic part is a $2-$form. The existence of all possible $2$-form
basis elements in the fermionic part is a reflection of its
generality.

\subsection{Matrix realization}

In terms of the $2\times 2$ Pauli matrices
\begin{eqnarray}
\sigma_{1}=\left(
\begin{array}{cc}
0 & 1 \\
1 & 0
\end{array}
\right),\quad \sigma_{2}=\left(
\begin{array}{cc}
0 & -i \\
i & 0
\end{array}
\right),\quad \sigma_{3}=\left(
\begin{array}{cc}
1 & 0 \\
0 & -1
\end{array}
\right)\;, \nonumber
\end{eqnarray}
and of the $2\times 2$ unit matrix $\textbf{1}$, we shall use
\begin{eqnarray}
e^{1}&=&\left(
\begin{array}{cc}
0 & i\sigma_{1} \\
-i\sigma_{1} & 0
\end{array}
\right),\;\;\;\quad e^{2}=\left(
\begin{array}{cc}
0 & i\sigma_{3} \\
-i\sigma_{3} & 0
\end{array}
\right),\nonumber\\
& & \\
 e^{3}&=&\left(
\begin{array}{cc}
0 & i\sigma_{2} \\
-i\sigma_{2} & 0
\end{array}
\right),\quad \quad e^{4}=\left(
\begin{array}{cc}
0 & \textbf{1} \\
\textbf{1} & 0
\end{array}
\right)\;,\nonumber
\end{eqnarray}
for the complex Clifford basis defined by (3). This is a
representation such that all basis matrices are Hermitian, while
$e^{3}$ and $e^{4}$ are symmetric and $e^{1}$ and $e^{2}$ are
antisymmetric. If the representations (24) are used in (22), we
obtain
\begin{equation}
H_{s} = \left(
\begin{array}{cc}
H_{1} & 0 \\
0 & H_{2}
\end{array}
\right)=H_{1}\pi_{+}+H_{2}\pi_{-} ,
\end{equation}
where $\pi_{+}=diag(\textbf{1},0)$ and $\pi_{-}=diag(0,\textbf{1})$
denotes the non-primitive projections and
\begin{eqnarray}
H_{1}=H_{\ast}\textbf{1} +H_{1F}\;,\quad H_{2}=H_{\ast}\textbf{1}
+H_{2F}\;.
\end{eqnarray}
$H_{jF}$'s represent the following fermionic parts
\begin{eqnarray}
H_{1F}&=&\frac{i}{2} B_{+}\sigma_{3}\;,\\
H_{2F}&=&\frac{i}{2}B_{-}\sigma_{3}-i[W_2,P_1]_M\sigma_1-i[W_1,W_2]_M\sigma_2\;,
\end{eqnarray}
with
\begin{eqnarray}
B_{\pm}=[W_{1},P_{1}]_{M}\pm [W_{2},P_{2}]_{M} \;.\nonumber
\end{eqnarray}
In deriving these relations we  made use of
$\sigma_{1}\sigma_{2}\sigma_{3}=i\textbf{1}$ and its consequences.
Note that $H_1$ is diagonal.

Having obtained matrix realizations of the Hamiltonians we should
emphasize that both of the $H_1$ and $H_2$ are Hermitian. Firstly,
in view of (14) $H_{\ast}$ given by (23) is a real-valued
phase-space function. Secondly, as $B_{\pm}$ and the other two
coefficient functions of $H_{2F}$ consisting Moyal brackets are
purely imaginary valued, the presence of the imaginary unit $i$ in
(27) and (28) ensures the Hermiticity of the both fermionic parts.

Equations (15) (or equivalently (17) and (18)), (20) and (22) show
the main differences between the Clifford algebra and the
Moyal-Clifford algebra. These restrictions given by (15) arise in
order to make $q_{\pm}$ nilpotent in the deformed case and play
important roles in the rest of this work. The resulting algebra is a
genuine SUSY algebra and the resulting Hamiltonians have
non-classical parts. The common bosonic part $H_{*}$ of $H_1$ and
$H_2$ has $H$ given by (5) as its classical limit:
$\lim_{\hbar\rightarrow 0}H_{*}=H$. However, their fermionic parts
are different and have no classical limits despite the fact that the
coefficient functions of the Pauli matrices have classical limits:
all Moyal brackets in (27) and (28) reduce to PB brackets in the
classical limit after dividing by $i\hbar$.

\section{Intertwining, Iso-spectral Property and Constants of Motion}

In terms of
\begin{eqnarray}
L_{1}&=& C_1(i\sigma_1+\sigma_2)-iC_2(\textbf{1}-\sigma_3)=2i\left(
\begin{array}{cc}
0 & 0\\
C_1 & -C_2
\end{array}
\right)\;,\\
L_{2}&=& C_1(i\sigma_1+\sigma_2)+iC_2(\textbf{1}+\sigma_3)=2i\left(
\begin{array}{cc}
C_2 & 0\\
C_1 & 0
\end{array}
\right)\;,
\end{eqnarray}
the matrix representations of the supercharges $q_{\pm}$ are found,
from (9) and (24), to be
\begin{equation}
q_+ =\frac{1}{\sqrt{2}} \left(
\begin{array}{cc}
0 & L_{1}\\
-L_{2} & 0
\end{array}
\right)= {q_-}^\dag\;.
\end{equation}
The matrices of $\omega_1$ and $\omega_2$ are Hermitian. Nilpotency
of $q_+$ implies that
\begin{eqnarray}
L_{1}*_{MC} L_2 =0= L_{2}*_{MC}L_{1}\;,
\end{eqnarray}
i.e. $L_1$ and $L_2$ are divisors of zero with respect to the
$*_{MC}$-product, which denotes, from now on, the star product of
matrix-valued phase-space functions. We should note that the second
equality of (32) results from the matrix product of factors
$i\sigma_1+\sigma_2$ and $(\textbf{1}\pm\sigma_3)$ appearing in (29)
and (30), while the condition (15) is essential for the first
equality of (32).

On the other hand, (20) implies the following double intertwining
relations:
\begin{eqnarray}
L_{2}*_{MC} H_1 &=& H_2*_{MC} L_{2}\;,\\
L_{1}*_{MC} H_2 &=& H_1*_{MC} L_{1}\;.
\end{eqnarray}
One can easily verify that $(D_{1}*_{MC} D_2)^{\dag} =
D^{\dag}_2*_{MC} D^{\dag}_{1}$ holds for arbitrary matrix valued
functions $D_j$'s. In view of this fact, the Hermitian conjugates of
(33) and (34), or equivalently the relation $[q_{-},H_s]_{MC}=0$,
gives the following additional intertwining relations
\begin{eqnarray}
L^{\dag}_{2}*_{MC} H_2 &=& H_1*_{MC} L^{\dag}_{2}\;,\\
L^{\dag}_{1}*_{MC} H_1 &=& H_2*_{MC} L^{\dag}_{1}\;.
\end{eqnarray}
As is evident from (32), $L^{\dag}_{1}$ and $L^{\dag}_{2}$ are also
divisors of zero. By virtue of (19) and (31), we also obtain
\begin{eqnarray}
H_1&=&\frac{1}{4}(L_{1}*_{MC}L^{\dag}_{1}+ L^{\dag}_{2}*_{MC}L_{2})\;,\\
H_2&=&\frac{1}{4}(L^{\dag}_{1}*_{MC}L_{1}+
L_{2}*_{MC}L^{\dag}_{2})\;,
\end{eqnarray}
which show that each partner Hamiltonian can be written as a sum of
factorized terms.

Let us call a nonzero $2\times 1$ matrix-valued function $\Psi$ a
star-eigenspinor of a $2\times 2$ matrix-valued function $T$
corresponding to the star-eigenvalue $\lambda$ if and only if
$T*_{MC} \Psi=\lambda \Psi$. This definition implies that
$\Psi^{\dag}*_{MC} T^{\dag}=\bar{\lambda}\Psi^{\dag}$, and that
$\Psi$ is nonzero if and only if $\Psi^{\dag}*_{MC}\Psi\neq 0$. Then
the standard theorems for Hermitian operators, in particular the
theorems concerning the reality of spectra and the orthogonality of
eigenfunctions corresponding to different eigenvalues, are also
valid in the present context. We use the term {\it spinor} in the
usual sense of $2\times 1$ matrix eigenfunction and remark that this
term has a wider meaning in the nomenclature of Clifford algebra.
Bearing these facts in mind, we now return to the physical
implications of the intertwining relations \cite{Kuru}.

The first important implication is the fact that $H_1$ and $H_2$ are
iso-spectral, that is, they have almost the same spectra. More
concretely, if $\Psi$ is an eigen-spinor of $H_1$ with eigenvalue
$\lambda$, then $L_{2}*_{MC}\Psi$ and $L_{1}^{\dag}*_{MC}\Psi$ are
also eigen-spinors of $H_2$ with the same eigenvalue provided that
$\Psi$ is not in the $MC$-kernel of $L_{2}$, or $L_{1}^{\dag}$. In
view of (34) and (35) the analogous remark is valid for the
eigen-spinors of $H_2$. For brevity, we refer to the following
section for more details and return to another implication of the
intertwining relations.

Multiplying one of the relations (33-36) by $L_{j}$ (or
$L^{\dag}_{j}$) from the left and comparing the resulting
expression with a similarly obtained one from the others, lead to
the following matrix valued functions:
\begin{eqnarray}
R_1&=&L_{1}*_{MC} L^{\dag}_1,\quad S_2=L^{\dag}_{2}*_{MC} L_2\;,\\
R_2&=&L_{2}*_{MC} L^{\dag}_2,\quad S_1=L^{\dag}_{1}*_{MC} L_1\;.
\end{eqnarray}
These commute with $H_1$ and $H_2$ in the following way:
\begin{eqnarray}
\;[R_1, H_1]_{MC}&=&0=[S_2, H_1]_{MC}\;,\\
\;[R_2,H_2]_{MC}&=&0=[S_1,H_2]_{MC}\;.
\end{eqnarray}
However they are not independent since $4H_1=R_1+S_2$ and $
4H_2=R_2+S_1$. That is, if there is no explicit time dependence,
each system has, together with the Hamiltonian, two constants of
motion. The explicit forms of the constants of motion for $H_1$ are
$S_{2}=2H_{1}(\textbf{1}+\sigma_3)$ and $R_{1}
=2H_{1}(\textbf{1}-\sigma_3)$. Since $H_1$ is diagonal, $S_2$ and
$R_1$ are its projected forms.

The method developed so far contains four real-valued phase-space
functions. Apart from conditions (17) and (18), there is no
constraint on these functions. Therefore, the $\ast$-products of
functions appearing in our formulae contain, in general, countably
infinite terms each characterized by a positive power of $\hbar$.
This generality enables us to study quantum properties of many
physically relevant systems on a classical phase space. However, as
we are about to do in the following section, for physically relevant
systems some of these functions should be restricted to finite
polynomial functions of the canonical coordinates. In such a case a
comparison with the existing literature of supersymmetric quantum
mechanics, in which majority of application have been carried out in
the usual operator formulation of quantum mechanics, can be made. In
this regard it is interesting to observe that the forms of our $H_s,
H_1$ and supercharges are similar, up to some permutations of rows
and columns of matrices, to that found in \cite{Ioffe}.

\section{Applications}

In terms of spin raising and lowering matrices
$\sigma_{\pm}=(\sigma_1\pm i\sigma_2)/2$, $H_{2F}$ given by (28) can
be rewritten as
\begin{eqnarray}
H_{2F}=\frac{i}{2}B_{-}\sigma_{3}+\sqrt{2}(\mathcal{A}\sigma_+
+\bar{\mathcal{A}}\sigma_-)\;,
\end{eqnarray}
where
\begin{eqnarray}
\mathcal{A}=[W_2,\bar{C}_1]_M\;,\quad \bar{\mathcal{A}}=-
[W_2,C_1]_M\;.
\end{eqnarray}
These are the phase-space analogues of the bosonic raising and
lowering operators when $[\mathcal{A},\bar{\mathcal{A}}]_M$ is a
(real) constant. To give illustrative examples we shall (from here
on) restrict the study to such cases, take $\hbar=1$ and denote the
second term of (43) as
\begin{eqnarray}
H_{JC}(\mathcal{A})=\sqrt{2}(\mathcal{A}\sigma_+
+\bar{\mathcal{A}}\sigma_-)\;,
\end{eqnarray}
when $[\mathcal{A},\bar{\mathcal{A}}]_M=1$. In such a case, $H_{JC}$
is the phase space analog of the well-known, fully quantum
mechanical Jaynes-Cummings (JC) Hamiltonian of quantum optics which
describes a bipartite (a two-level atom and a field) system. In the
parlance of quantum optics, $\sigma_{\pm}$ are known as the atomic
transition operators and $\sigma_3$ as the inversion operator. The
first $\mathcal{A}\sigma_{+}$ term at the right-hand side of (45)
represents an absorption process in which destruction of a field
quanta and a transition to a higher atomic level take place. The
second term $\bar{\mathcal{A}}\sigma_{-}$ describes an emission
process in which creation of a field quanta is accompanied by a
transition to a lower atomic level.

Let us see how the phase-space version of such a model together with
its supersymmetric partner naturally emerges from the method
developed above.

\subsection{Example 1: Jaynes-Cummings-type systems}

A particular simple choice of $W_j$ and $P_1$ amenable to various
physical applications seems to be as
\begin{eqnarray}
W_1=p_2\;,\quad P_1=q_2\;,\quad W_2=q_1p_2-q_2p_1\;.
\end{eqnarray}
Then, in view of $[W_1,W_2]_M=ip_1$ and $[W_2,P_1]_M=-iq_1$, the
conditions (17) and (18) imply that the most general form of $P_2$
is
\begin{eqnarray}
P_2=p_1p_2+q_1q_2+K_{1}\;,
\end{eqnarray}
such that $K_{1}=K_{1}(q_1,p_1)$. It is now straightforward to
verify, with $C_1=(p_2+iq_2)/\sqrt{2}$ from (44), that
\begin{eqnarray}
\mathcal{A}&=&-\frac{1}{\sqrt{2}}(q_1+ip_1)\;, \quad
[\mathcal{A},\bar{\mathcal{A}}]_{M}=1\;.
\end{eqnarray}
Hence, $\mathcal{A}$ and $\bar{\mathcal{A}}$ are bosonic lowering
(annihilation) and raising (creation) phase-space functions,
respectively. One can also define another such bosonic pair by
\begin{eqnarray}
\mathcal{B}=-\frac{1}{\sqrt{2}}(q_2+ip_2)\;, \quad
[\mathcal{B},\bar{\mathcal{B}}]_{M}=1\;,
\end{eqnarray}
which Moyal commute with the above pair. $W_2$ represents the
angular momentum perpendicular to the $q_1q_2$-plane and, together
with $P_2$, can be rewritten as
\begin{eqnarray}
W_2=-i(\bar{\mathcal{A}}\mathcal{B}-\mathcal{A}\bar{\mathcal{B}})\;,\quad
P_2=\mathcal{A}\bar{\mathcal{B}}+\bar{\mathcal{A}}\mathcal{B}+K_{1}\;,
\end{eqnarray}

Let us define the phase-space number functions
\begin{eqnarray}
N_{\mathcal{A}}=\bar{\mathcal{A}}\ast\mathcal{A}=\bar{\mathcal{A}}\mathcal{A}
-\frac{1}{2},\;\;N_{\mathcal{B}}=
\bar{\mathcal{B}}\ast\mathcal{B}=\bar{\mathcal{B}}\mathcal{B}
-\frac{1}{2}\;.
\end{eqnarray}
By making use of
\begin{eqnarray}
F_1({\bf q})\ast F_{2}({\bf q})=F_{1}({\bf q})F_{2}({\bf q})\;,\quad
G_{1}({\bf p})\ast G_{2}({\bf p})= G_{1}({\bf p}) G_{2}({\bf
p})\;,\nonumber
\end{eqnarray}
and $G\ast_{M}G=G^{2}$ for $G={\bf a}\cdot{\bf p}+F_1({\bf q})$
(${\bf a}$ is a constant vector), we also find
\begin{eqnarray}
H_{2F}&=&\frac{i}{2}B_{-}\sigma_{3}+H_{JC}(\mathcal{A})\;,\nonumber\\
B_{\pm}&=&i\big[-1\pm 2(N_{\mathcal{B}}-N_{\mathcal{A}}+X_{1})\big]\;,\\
H_{\ast}&=&H_{\mathcal{A}}+2N_{\mathcal{B}}(N_{\mathcal{A}}+1)+
Y_{1}\;,\nonumber
\end{eqnarray}
where $H_{\mathcal{A}}=(2N_{\mathcal{A}}+1)/2$, and
\begin{eqnarray}
X_{1}&=&-\frac{i}{2}[W_2,K_{1}]_M\;,\\
Y_{1}&=&\frac{1}{2}\{\mathcal{A}\bar{\mathcal{B}}+\bar{\mathcal{A}}\mathcal{B},K_{1}\}_M+
\frac{1}{2}K_{1}\ast K_{1}\;.
\end{eqnarray}
$H_2$ is of the type of a JC Hamiltonian \cite{JC1,JC2,Schleich}
describing a two-level atom interacting with a quantized, two-mode
electromagnetic field of which only the $\mathcal{A}$-mode
directly interacts with the atom and hence causing transitions
between levels. $H_{\mathcal{A}}$ in $H_{\ast}$ is the energy of
the $\mathcal{A}$-mode and the remaining part
\begin{eqnarray}
H_{int}=H_{\ast}-H_{\mathcal{A}}=2N_{\mathcal{B}}(N_{\mathcal{A}}+1)+
Y_{1}\;,\nonumber
\end{eqnarray}
represents the energy of the other mode as well as the interaction
between the modes. The super-partners are
\begin{eqnarray}
H_{1}&=&(H_{int}+H_{\mathcal{A}})\textbf{1}+(H_{\mathcal{A}}-N_{\mathcal{B}})\sigma_3\;,\\
H_{2}&=&(H_{int}+H_{\mathcal{A}})\textbf{1}+(H_{\mathcal{B}}-N_{\mathcal{A}})\sigma_3
+H_{JC}(\mathcal{A}) \;,
\end{eqnarray}
where $H_{\mathcal{B}}=(2N_{\mathcal{B}}+1)/2$, we have taken
$K_{1}=0$ and hence $H_{int}=2N_{\mathcal{B}}(N_{\mathcal{A}}+1)$.
In such a case the interactions between the modes are through the
number functions. By choosing $K_{1}$ such that $X_{1}$ and $Y_{1}$
are nonzero, one can also model certain types of mode interactions
\cite{Barnett}. Even in the simplest case when $K_{1}$ is a nonzero
constant, $X_{1}$ is zero but the $Y_{1}$ term, which becomes
$K_1(\mathcal{A}\bar{\mathcal{B}}+\bar{\mathcal{A}}\mathcal{B})+(K_1^{2}/2)$,
represents a coherent exchange of photons between the modes. In any
case $H_1$ has no JC-type atom-field interaction term.

\subsection{Eigenvalues, Eigenspinors and Wigner Functions}

$H_1$ is diagonal and algebraically depends on the complete set
$\{N_{\mathcal{A}}\textbf{1},\;N_{\mathcal{B}}\textbf{1},\;\sigma_3\}$
of three mutually commuting (in the MC sense) matrix-valued
phase-space functions. Using the shortened notation
$(\textbf{z}|\dots)=(\textbf{q},\textbf{p}|\dots)$, their common
eigenspinors will be denoted by
\begin{equation}
(\textbf{z}|j,n_{\mathcal{A}},n_{\mathcal{B}})=|j\rangle\otimes
(\textbf{z}|n_{\mathcal{A}},n_{\mathcal{B}})\;,
\end{equation}
Here $n_{\mathcal{A}},n_{\mathcal{B}}=0,1,2,\dots$, stand for the
numbers of mode-quanta and $j(=1,2)$ labels the bare states of atom
such that $|j\rangle$'s, like the spin-up and spin-down states of a
spin-$1/2$ system, satisfy $\sigma_3|1\rangle=-|1\rangle,
\sigma_3|2\rangle=|2\rangle$ and
\begin{eqnarray}
\sigma_-|1\rangle &=& 0=\sigma_+|2\rangle\;,\nonumber\\
\sigma_-|2\rangle &=& |1\rangle\;,\nonumber\\
\sigma_+|1\rangle &=& |2\rangle\;.\nonumber
\end{eqnarray}
In (57) the real-valued functions
$(\textbf{z}|n_{\mathcal{A}},n_{\mathcal{B}})$ represent the
diagonal Wigner functions \cite{Fairly,Curtright}
\begin{eqnarray}
 (\textbf{z}|n_{\mathcal{A}},n_{\mathcal{B}})=\frac{1}
{n_{\mathcal{A}}!n_{\mathcal{B}}!}
\bar{\mathcal{A}}^{n_{\mathcal{A}}}\ast_{M}\bar{\mathcal{B}}^{n_{\mathcal{B}}}\ast_{M}
(\textbf{z}|0,0)\ast_M\mathcal{A}^{n_{\mathcal{A}}}\ast_{M}\mathcal{B}^{n_{\mathcal{B}}}\;,
\end{eqnarray}
for two-mode field. Here $(\textbf{z}|0,0)$ denotes the vacuum
Wigner function defined by
\begin{eqnarray}
\mathcal{A}\ast_{M}\langle\textbf{z}|0,0\rangle=0=\mathcal{B}\ast_{M}\langle\textbf{z}|0,0\rangle
\;.\nonumber
\end{eqnarray}
The normalized (its integral all over the phase space set to $1$)
solution of these equations is
$\pi^{-2}\exp-(p_1^{2}+p_1^{2}+q_1^{2}+q_1^{2})$. Explicit
functional forms of the higher level Wigner functions are
proportional to \cite{bengu}
\begin{eqnarray}
L_{n_{\mathcal{A}}}(4\mathcal{A}\bar{\mathcal{A}})
L_{n_{\mathcal{B}}}(4\mathcal{B}\bar{\mathcal{B}})
\langle\textbf{z}|0,0\rangle\;,\nonumber
\end{eqnarray}
where $L_{k}$ stands for the Laguerre polynomials. These are
products of the harmonic oscillator Wigner functions which are well
known \cite{Moyal}.

One can also verify
\begin{eqnarray}
\mathcal{A}\ast_{M} (\textbf{z}|n_{\mathcal{A}},n_{\mathcal{B}})&=&
\sqrt{n_{\mathcal{A}}}\;(\textbf{z}|n_{\mathcal{A}}-1,n_{\mathcal{B}})\;,\nonumber\\
\bar{\mathcal{A}}\ast_{M}(\textbf{z}|n_{\mathcal{A}},n_{\mathcal{B}})&=&
\sqrt{n_{\mathcal{A}}+1}\;(\textbf{z}|n_{\mathcal{A}}+1,n_{\mathcal{B}})\;,\\
N_{\mathcal{A}}\ast_{M}
(\textbf{z}|n_{\mathcal{A}},n_{\mathcal{B}})&=&
n_{\mathcal{A}}\;(\textbf{z}|n_{\mathcal{A}},n_{\mathcal{B}})\;.\nonumber
\end{eqnarray}
Similar relations hold for $N_{\mathcal{B}},\mathcal{B}$ and
$\bar{\mathcal{B}}$. In view of these relations, the eigenvalues of
$H_1$
\begin{eqnarray}
H_1\ast_{MC} (\textbf{z}|1,
n_{\mathcal{A}},n_{\mathcal{B}})&=&\lambda_{1n_{\mathcal{A}}n_{\mathcal{B}}}(\textbf{z}|1,
n_{\mathcal{A}},n_{\mathcal{B}})\;,\nonumber \\
H_1\ast_{MC} (\textbf{z}|2,
n_{\mathcal{A}},n_{\mathcal{B}})&=&\lambda_{2n_{\mathcal{A}}n_{\mathcal{B}}}(\textbf{z}|1,
n_{\mathcal{A}},n_{\mathcal{B}})\;,\nonumber
\end{eqnarray}
are easily found, from (55) and (59), to be
\begin{eqnarray}
\lambda_{1n_{\mathcal{A}}n_{\mathcal{B}}}=n_{\mathcal{B}}(2n_{\mathcal{A}}+3)
\;,\quad
\lambda_{2n_{\mathcal{A}}n_{\mathcal{B}}}=(n_{\mathcal{B}}+1)(2n_{\mathcal{A}}+1)\;.
\end{eqnarray}
For $n_{\mathcal{A}}=0=n_{\mathcal{B}}$, we have $\lambda_{100}=0$
and $\lambda_{200}=1$ which represent the bare energy levels of the
atom. In this case, $H_1$ is simply $(\textbf{1}+\sigma_3)/2$. The
level $\lambda_{1n_{\mathcal{A}}0}=0$ is infinitely degenerate
whereas all the other levels are finitely degenerate. Provided that
$n_{\mathcal{A}}\geq n_{\mathcal{B}}$, the level
$\lambda_{2n_{\mathcal{A}}n_{\mathcal{B}}}$ is the higher level of
the whole atom-field system. These degeneracies arise from the fact
that we have taken the frequencies of the modes to be equal
(degenerate modes). Evidently, these discussions can be extended to
general case in which the frequencies are different.

For this example, we have
$C_2=i\sqrt{2}\bar{\mathcal{B}}\mathcal{A},\;C_1=-i\bar{\mathcal{B}}$
and by virtue of
\begin{eqnarray}
2\sigma_{+}\sigma_{-}=\textbf{1}+\sigma_3\;,\quad
2\sigma_{-}\sigma_{+}=\textbf{1}-\sigma_3\;,
\end{eqnarray}
the intertwining (matrix-valued) functions, given by (29) and (30)
can be rewritten as
\begin{eqnarray}
L_1&=&2\bar{\mathcal{B}}\sigma_{-}(\textbf{1}+\sqrt{2}\mathcal{A}\sigma_+)\;,\quad
L^{\dag}_1=2\mathcal{B}(\textbf{1}+\sqrt{2}\bar{\mathcal{A}}\sigma_-)\sigma_+\;,\\
L_2&=&2\bar{\mathcal{B}}(\textbf{1}-\sqrt{2}\mathcal{A}\sigma_{+})\sigma_{-}\;,\quad
L^{\dag}_2=2\mathcal{B}\sigma_+(\textbf{1}-\sqrt{2}\bar{\mathcal{A}}\sigma_-)\;.
\end{eqnarray}
Now it is easy to verify
\begin{eqnarray}
L_2\ast_{MC}(\textbf{z}|1,n_{\mathcal{A}},n_{\mathcal{B}})&=&0=
L^{\dag}_1\ast_{MC}(\textbf{z}|2,n_{\mathcal{A}},n_{\mathcal{B}}),\nonumber
\end{eqnarray}
which simply follow from the actions of $\sigma_{\pm}$ on the atomic
bare states. However, one needs to be more careful for the
verifications of
\begin{eqnarray}
L^{\dag}_1\ast_{MC}(\textbf{z}|1,n_{\mathcal{A}},n_{\mathcal{B}})&=&
2\sqrt{n_{\mathcal{B}}}\;\Phi_{n_{\mathcal{A}}n_{\mathcal{B}}}(\textbf{z})\;,\nonumber\\
L_2\ast_{MC}(\textbf{z}|2,n_{\mathcal{A}},n_{\mathcal{B}})&=&
2\sqrt{n_{\mathcal{B}}+1}\;\Psi_{n_{\mathcal{A}}n_{\mathcal{B}}}(\textbf{z})\;,\nonumber
\end{eqnarray}
where
\begin{eqnarray}
\Phi_{n_{\mathcal{A}}n_{\mathcal{B}}}(\textbf{z})&=&\sqrt{2n_{\mathcal{B}}+2}\;
(\textbf{z}|1,n_{\mathcal{A}}+1,n_{\mathcal{B}}-1)+(\textbf{z}|2,n_{\mathcal{A}},n_{\mathcal{B}}-1)\;,\\
\Psi_{n_{\mathcal{A}}n_{\mathcal{B}}}(\textbf{z})&=&
(\textbf{z}|1,n_{\mathcal{A}},n_{\mathcal{B}}+1)-\sqrt{2n_{\mathcal{A}}}\;
(\textbf{z}|2,n_{\mathcal{A}}-1,n_{\mathcal{B}}+1).
\end{eqnarray}
As a consistency check, one can directly verify that these are the
(unnormalized) eigen-spinors of $H_2$
\begin{eqnarray}
H_2\ast_{MC}\Phi_{n_{\mathcal{A}}n_{\mathcal{B}}}(\textbf{z})&=&
\lambda_{1n_{\mathcal{A}}n_{\mathcal{B}}}\Phi_{n_{\mathcal{A}}n_{\mathcal{B}}}(\textbf{z})\;,\\
H_2\ast_{MC} \Psi_{n_{\mathcal{A}}n_{\mathcal{B}}}(\textbf{z})&=&
\lambda_{2n_{\mathcal{A}}n_{\mathcal{B}}}\Psi_{n_{\mathcal{A}}n_{\mathcal{B}}}(\textbf{z})\;,
\end{eqnarray}
with the same eigenvalues as given by (60). These fully establish
the iso-spectral property of $H_1$ and $H_2$, which was compactly
expressed by the SUSY algebra given by (19-21).

In the nomenclature of quantum optics the $\ast$-eigenstates given
by (64) and (65) are the phase-space analogues of the so-called
stationary \textit{dressed states} or \textit{JC-doublet}. The
states at the right-hand sides of (64) and (65) are known as the
bare states of the atom-field system. The latter are the phase-space
version of the product states of the bare atom and field states.
Finally, we compute, with the help of (62) and (63), the constants
of motion
\begin{eqnarray}
R_1&=&8N_{\mathcal{B}}(H_{\mathcal{A}}+1)\sigma_{-}\sigma_{+}\;,\\
S_1&=&4(N_{\mathcal{B}}+1)[\sigma_{+}\sigma_{-}+2N_{\mathcal{B}}\sigma_{-}\sigma_{+}+
H_{JC}(\mathcal{A})]\;,
\end{eqnarray}
such that $[R_1,H_1]_{MC}=0=[S_1,H_2]_{MC}$.

\subsection{Example 2: Non-resonant JC-type interactions}

For $W_1=p_1,\;P_1=-q_1$ and $W_2$ as in (46), the conditions (17)
and (18) are satisfied, provided that
\begin{equation}
P_2=p_1p_2+q_1q_2+K_2\;.\nonumber
\end{equation}
Here $K_2=K_2(q_2,p_2)$ is an arbitrary real function of its
arguments. Hence $\sqrt{2}C_1=p_1-iq_1$ and
\begin{equation}
[W_2,\bar{C_1}]_M=-(q_2-ip_2)/\sqrt{2}=\bar{\mathcal{B}}\;,\nonumber
\end{equation}
where $\mathcal{B}$ and $\bar{\mathcal{B}}$ are defined by (49). As
a result, by interchanging the lowering functions of subsection
\textbf{V}.\textbf{A} such that
$(\mathcal{A},\mathcal{B})\leftrightarrow
(\bar{\mathcal{B}},\bar{\mathcal{A}})$, or by direct computations,
we find
\begin{eqnarray}
H_{2F}&=&\frac{i}{2}B_{-}\sigma_{3}+H_{JC}(\bar{\mathcal{B}})\;,\nonumber\\
H_{JC}(\bar{\mathcal{B}})&=&\sqrt{2}(\bar{\mathcal{B}}\sigma_+
+\mathcal{B}\sigma_-)\;,\nonumber\\
B_{\pm}&=&i\big[1\pm 2(N_{\mathcal{B}}-N_{\mathcal{A}}+X_{2})\big]\;,\nonumber\\
H_{\ast}&=&H_{\mathcal{B}}+2N_{\mathcal{A}}(N_{\mathcal{B}}+1)+
Y_{2}\;.\nonumber
\end{eqnarray}
$W_2$ and the first term of $P_2$ are still given by (50), and
$X_2,Y_2$ are defined as in (53) and (54) provided that $K_1$ is
replaced by $K_2$. For $K_2=0$, the Hamiltonians and the
corresponding eigenvalues are
\begin{eqnarray}
H_{1}&=&(H_{int}+H_{\mathcal{B}})\textbf{1}-(H_{\mathcal{B}}-N_{\mathcal{A}})\sigma_3\;,\nonumber\\
H_{2}&=&(H_{int}+H_{\mathcal{A}})\textbf{1}-(H_{\mathcal{A}}-N_{\mathcal{B}})\sigma_3
+H_{JC}(\bar{\mathcal{B}}),\nonumber\\
\lambda_{1n_{\mathcal{A}}n_{\mathcal{B}}}&=&(n_{\mathcal{A}}+1)(2n_{\mathcal{B}}+1),\;
\lambda_{2n_{\mathcal{A}}n_{\mathcal{B}}}=n_{\mathcal{A}}(2n_{\mathcal{B}}+1).\nonumber
\end{eqnarray}
These are related to (60) by the interchange
$(1,n_{\mathcal{A}},n_{\mathcal{B}})\leftrightarrow
(2,n_{\mathcal{B}},n_{\mathcal{A}})$.

Now some remarks for the above applications are in order.

The JC-type Hamiltonians originate from the interaction of the
electric dipole moment of an atom with a quantized light in dipole
approximation and, in general, both $H_{JC}(\mathcal{A})$ and
$H_{JC}(\bar{\mathcal{A}})$ take part in such an interaction. To see
this we should emphasize that dipole moment has non-vanishing matrix
elements only between states of opposite parity. Therefore, the
atomic bare states $|1\rangle$ and $|2\rangle$ are assumed of
opposite parity and hence the dipole moment is proportional to
$\sigma_+ +\sigma_-=|1\rangle\langle 2|+|2\rangle\langle 1|$. On the
other hand since a quantized single mode cavity field is a multiple
of $\mathcal{A}+\bar{\mathcal{A}}$ then the dipole energy is
proportional to
\begin{eqnarray}
(\sigma_++\sigma_-)(\mathcal{A}+\bar{\mathcal{A}})=
H_{JC}(\mathcal{A})+H_{JC}(\bar{\mathcal{A}})\;.\nonumber
\end{eqnarray}
However, the resonant processes described by $H_{JC}(\mathcal{A})$
are more efficient than the non-resonant processes described by
$H_{JC}(\bar{\mathcal{A}})$, especially in the case of a quantized
single-mode light. Neglect of the latter is usually called the {\it
rotating wave approximation}. On the other hand, in the case of
multi-mode interactions, both types of terms have important physical
implications. Here we will be content with pointing out that the
above two examples can be combined and then extended in various ways
to study the generalized models of quantum optics, such as $N$-atom
JC models \cite{Chagas} (also known as Dick models \cite{Brandes1})
and spin-boson systems \cite{Brandes2}. In particular, time can be
included and the dynamics of transitions between atomic levels for
both partner Hamiltonians can be studied in the phase space.

\section{Concluding Remarks}

The Moyal-Clifford algebra, that is, Clifford algebra endowed with
the star product of deformation quantization, provides a unified
framework to realize the SUSY techniques in a classical phase space,
to see relations of various models described by matrix Hamiltonians
and to study their phase-space characteristics in great detail. To
emphasize some merits of such an approach we recall that in the
usual operator formulation of quantum mechanics, Wigner functions
are indirectly obtained from wave functions through some
\textit{convolution} integrals which, apart from simple few cases,
are not easy to cope with. But, in the deformation quantization they
are directly obtained from star-eigenvalue equations. Moreover, by
realizing the SUSY techniques in this framework Wigner functions of
some delicate states such as the dressed states can be easily
obtained by the intertwining functions.

In the presented examples the phase space is spanned, instead of the
usual canonically conjugate coordinates of a particle system, by the
conjugate amplitudes of mode functions of a quantized
electromagnetic field that are dynamically equivalent to conjugate
coordinates of a mechanical oscillator. The form of $H_1$ and $H_2$
and the presented examples may be convincing enough that a lot of
physically relevant (charged or uncharged) particle systems can be
identified as spacial cases. As much as exhaustive search in this
direction seems to be rather involved and will be deferred to a
later publication.

\begin{acknowledgments}
We are grateful to anonymous referees for their useful comments and
to \"{O}. Sar\i o\u{g}lu for a critical reading of the manuscript
and for illuminating discussions. Special thanks are due to A. U.
Y\i lmazer for useful conversations.  This work was supported in
part by the Scientific and Technological Research Council of Turkey
(T\"{U}B\.{I}TAK).
\end{acknowledgments}

\end{document}